\documentclass[12pt]{iopart}

\usepackage{graphicx,color}
\def\be{\begin{eqnarray}}
\def\ee{\end{eqnarray}}
\def\del{\partial}

\begin{document}

\title[Mach Cones in Quark Gluon Plasma]{Mach Cones in Quark Gluon Plasma}

\author{J Casalderrey-Solana}

\address{Lawrence Berkeley National Laboratory, 1 Cyclotron Road, MS 70R0319, Berkeley, CA 94720, USA  }
\ead{JCasalderrey-Solana@lbl.gov}
\begin{abstract}
The experimental azimuthal dihadron distributions at RHIC show a  double peak structure
in the away side ($\Delta \phi = \pi \pm 1.2$ rad.) 
 for intermediate $p_t$ particles. A variety of models have appeared trying to describe
this modification. We will review most of them, with special emphasis in the Conical Flow scenario in 
which the observed shape is a consequence of the emission of sound by a supersonic high momentum particle
propagating in the Quark Gluon Plasma.

\end{abstract}

%\maketitle

\section{Introduction}
One of the major findings of the relativistic heavy ion program
at RHIC is jet quenching 
\cite{jetquenching1,jetquenching2}, 
the suppression of high momentum particles. 
This suppression is a consequence of the energy loss of partons when
they traverse the hot and dense matter formed in a Au-Au collision. 
Even though at high energy the  loss 
is dominated by radiative processes \cite{Gyulassy_losses,Kovner:2003zj,Dok_etal},
 at intermediate momentum the 
microscopic mechanism for this process is not clear and there may 
be significant contribution from collisional processes as well \cite{Wicks:2005gt}. 

From the point of view of the fate of the energy lost by the probe,
the different scenarios can be classified in two major categories:
\begin{itemize}
\item Those in which the  energy  is deposited in the medium, 
such as collisional losses 
or absorption of the radiated gluons. The good hydrodynamical
behavior of the medium leads to collective effects, {\it i. e.},  the formation of a Mach Cone  
\cite{Stocker,Casalderrey-Solana:2006sq,Casalderrey-Solana:2004qm}.
\item  Those in which the energy is transferred to 
propagating modes that leave the interaction region.  These may be gluons
with dispersion relation close to the vacuum ones \cite{Vitev:2005yg,Polosa:2006hb}, 
or plasma modes with
significantly  modified dispersion relations (Cherenkov \cite{Koch:2005sx,Majumder:2005sw,Dremin:2005an} and
 Plasmon emission \cite{Ruppert:2005uz}).
\end{itemize}
The redistribution of the jet energy and momentum is reflected in the correlations
of particles associated with the jet. In fact, the experimental dihadron
correlation function shows, at intermediate $p_t$,
a double peak structure in the away side with the maximum of the 
correlation at $\Delta \phi \approx \pi \pm 1.2 $ rad. \cite{phenix_peaks,star_peaks}.
 The study of 
three particle correlations also indicates that the structure responsible
for this modification is conical \cite{threePPhenix,threePStar}. 

\begin{figure}
\begin{center}
%  \hspace{1cm}
 \includegraphics[width=7cm]{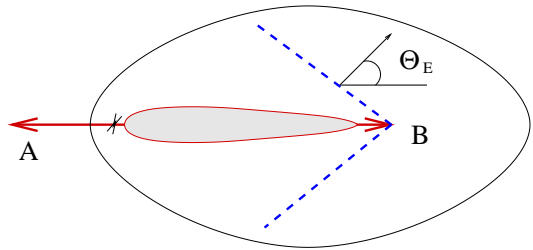}
 \caption{\label{fig_shocks}
A schematic picture of the relevant geometry for the correlation
measurements.
The trigger jet (A) travels a short distance in the medium while the
backjet (B) propagates though the entire medium. 
The interaction with the medium leads to emission of particles at
an angle $\theta_E\approx 1.2$ rad from the back jet. In the
 Mach cone scenario 
the dashed line corresponds to the Mach front, at an angle $\theta_M$
(\ref{MachAngle}), and the dotted region to the diffuson mode. A similar picture 
describes the Cherenkov and large angle gluon radiation scenarios.
}
\end{center}
\end{figure}

The experimental situation of these measurements 
 is sketched in Figure \ref{fig_shocks}. 
 The strong quenching biases the 
observed events to be 
produced close to the surface of the interaction region so that  
the path through matter of the trigger jet (A) is minimal. In turn, the back jet (B) 
travels a long distance through the medium and looses significant 
part of (if not all) its energy. Thus, the correlations
in the away side are strongly modified, either because of the 
formation of a Mach like front, of because of the medium induced 
gluon radiation.  
 
The observed structure of the dihadron distribution may be due, instead, to the
deflection of the parent parton \cite{Chiu:2006pu}
or the jet shower \cite{Armesto:2004pt} due to the medium, as 
opposed to a redistribution of the energy and momentum of the jet, what leads 
to a non conical particle distribution.

In what is next we will discuss the different models that attempt to 
explain the observed large angle correlations.

\section{Medium Absorption of the Lost Energy.}
\subsection{Hydrodynamic Description.}
Under the assumption that a significant part of the lost energy-momentum is absorbed and 
thermalized, we can describe its evolution 
 by means of hydrodynamics. The 
energy-momentum conservation equation is modified as,
\be
\del_{\mu} T^{\mu \nu}= J^{\mu} \,,
\ee
where the source $J$ codifies the loss process as well as the subsequent relaxation 
of the initial (non-hydrodynamical) disturbance.
Even though the total energy and
momentum lost 
 constraints the value of $J$, such constraint is not enough to determine it uniquely.

An accurate description of the backreaction of the medium requires the functional form  
of $J$. Unfortunately, the lack of knowledge about the interaction and thermalization
process of the energy lost prevents us from knowing these details. 
 However, within the context of linearized hydrodynamics in a static medium, a systematic study of the hydrodynamic
fields was performed for a general source \cite{Casalderrey-Solana:2006sq}. 
The basic finding is that there are two possible excitation modes
\begin{itemize}
\item Sound, a propagating mode. The interference of sound waves from a supersonic source leads to 
the Mach Cone, a conical flow  directed at an angle from the jet \be \label{MachAngle} \cos\theta_M=c_s \ee
\item Diffuson, a non propagating mode. This disturbance remains in the deposition region and leads to a
flow in the direction of the energetic particle. 
\end{itemize}

It was also shown that the production of entropy  
controls the different excitations \cite{Casalderrey-Solana:2006sq}. 
The amplitude of the diffuson mode is proportional to the amount of entropy produced in the particle-fluid interaction and the subsequent 
themalization. This motivated the introduction of two types of excitations: {\bf Isentropic excitation},
in which no significant entropy is produced and only  sound waves are emitted; {\bf no isentropic excitation},
 where significant entropy is produced and the diffuson mode is excited.

The two kinds of excitations have different manifestations in the 
dihadron correlations  \cite{Casalderrey-Solana:2006sq}. The spectrum of 
soft particles after the jet passage is given by the Cooper Fry prescription
(for the  static fluid we have equal time freeze out
conditions).
\be
\label{mspectrum}
\frac{dN}{d^3p}=\int_V \frac{d^3V}{2\pi^3} e^{-\frac{E}{T}+\delta} \, ,
\hspace{1.0cm}
\delta= \frac{E}{T} \frac{\delta T}{T} + \frac{{\bf p}{ \bf v}}{T} \, .
\ee
The angular 
distribution of low momentum particles
($p \sim T$) does not reflect the details of the flow picture due to the 
large thermal broadening. However, the high momentum ones do reflect the 
flow and, thus, their angular distribution is different for each
excitation mechanism. In the non isentropic case, the particles are mainly produced
in the direction of the probe. On the contrary, in the isentropic case
the flow moves along the Mach angle and the spectrum reflect this fact.

\begin{figure}[t]
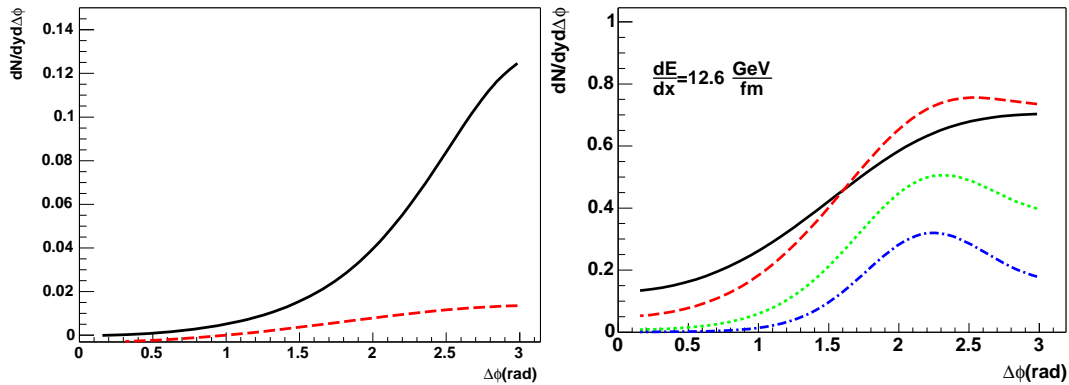

%\begin{center}
\includegraphics[width=5cm,angle=-90]{dNdydphiNI.epsi}
\includegraphics[width=5cm,angle=-90]{dNdydphipt_large.epsi}
%\end{center}
\begin{center}
\caption{
\label{corptdep}
Two particle correlation for the different hydrodynamical medium excitations. 
Left: non isentropic excitations, with $10 < p^{assoc}_t/T < 25$ and 
$dE/dx=63\, T^2$ (solid) and $dE/dx= 10 \, T^2$ (dashed).
Right: isentropic excitations for $dE/dx=63 \,T^2$ for 
$1 < p^{assoc}_t/T< 5$ (solid),
$5 < p^{assoc}_t/T< 10$ (dashed),
$10 < p^{assoc}_t/T< 15$ (dotted),
$15 < p^{assoc}_t/T< 20$ (dash-dotted).
The quoted value of $dE/dx$ corresponds to $T=200$ MeV \cite{Casalderrey-Solana:2006sq}.
%Isentropic Excitations.Associate yield dependence on associate $p_T$ 
%for fixed source size $\sigma=0.75/T$
% , viscosity $\Gamma_s=0.1/T$, $t_j=8/T$, $t_f=10/T$,
%and energy loss, $dE/dx=10 T^2$ (left) and $dE/dx=63 T^2$ 
%(right). The label values for $dE/dx$ correspond to 
%$T=200~\mbox{MeV}$.
% The three curves are for  
% $1 T<p_{t}<5 T$ (solid),
% $5 T<p_{t}<10 T$ (dotted),
%($3\times$) $10 T<p_{t}<15 T$ (dashed),
%($10\times$) $15 T<p_{t}<20 T$ (dashed-dotted).
%(in the upper panel all the curves are rescaled further up
%by a factor 10). No large angle correlation is observed
%for $dE/dx=10 T^2$. For $dE/dx=63 T^2$
%the position of the peak shifts toward $\pi$
%for lower $p_T$.
}
\end{center}
\end{figure} 

This behavior of the correlation function can be observed in 
Figure \ref{corptdep}. In Figure \ref{corptdep} a) the correlation function
for non isentropic excitations is plotted and, as a consequence of the diffuson
mode, the correlation is concentrated along the jet direction.
Similar findings were obtained in \cite{Heinz,Chaudhuri:2006qk} where non linear hydrodynamics was solved.
 In Figure \ref{corptdep}
b) the correlation function for the isentropic case is studied at different $p_t$ intervals.
As explained above, the low momentum part does not reflect the Mach cone, but as the associate 
momentum is increased the angle of maximum correlation is shifted off $\pi$ and saturates at
a value given by the  Mach angle. However, in order to get significant correlation strength 
at realistic values of the associate $p_t$ in this static approximation, very large values of the 
energy loss (and jet energy) are required.

This large value of the energy loss is an artifact of the static approximation. In a dynamic medium,
since the density of the fluid where the perturbation propagates decreases, the effective amplitude of the sound waves
increases
and one can get larger correlation strength for realistic values of the jet energy loss \cite{CS}.
The expansion also affects the direction of propagation of the shock. As the RHIC fireball cools down
the speed of sound of matter changes from $c^2_s=1/3$ in the QGP phase to $c_s\approx 0$ in the mixed 
phase and to $c^2_s\approx 0.2$ in the hadron gas \cite{Hung:1997du,Welke:1990za}. 
Taking from hydrodynamical simulations \cite{Teaney:2001av,Kolb:2001qz} the duration
of these three phases, the peak position is give by
\be
cos\theta_M=\bar  c_s={1/\tau}\int_o^\tau dt c_s(t)\approx 0.3
\Longrightarrow \Delta \phi \approx \pi- 1.2 \,,
\ee
which agrees with the experimental observations. 

Since the sound waves move
with the speed of sound in the fluid rest frame and due to the strong transverse 
flow of the RHIC fireball, one expects a broadening in the typical direction of the fluid with 
respect to $\theta_M$ \cite{S_w}. The 
strong longitudinal flow also leads to a $\eta$ broadening of the signal. 

\begin{figure}[t]
\begin{center}
\includegraphics[width=7cm]{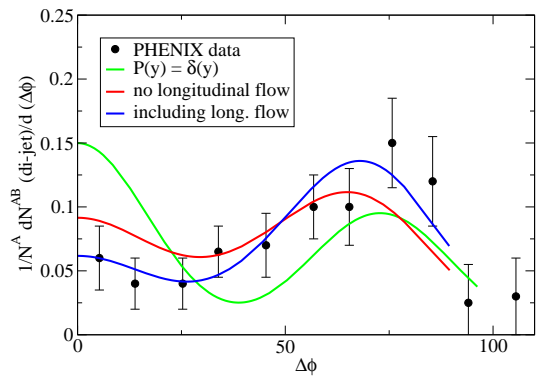}
\caption{
\label{RenkRuppert}
Two particle correlations computed by Renk and Ruppert in \cite{Renk:2006mv} for
different cases: a) fixed rapidity of the back jet and no coupling to 
flow (green) b) smeared back jet and no coupling to flow (red) and c)
smeared back jet and coupling to flow.
}
\end{center}
\end{figure}

These issues were addressed in \cite{Renk,Renk:2006mv}. In this work, the authors  do not solve hydrodynamics but 
introduced a simplified model that captures the main physics of the hydrodynamical solution. 
A fraction of energy $f=0.75$ is transferred into the sound mode, which propagates
at the Mach angle. The remaining energy $1-f$ propagates in the jet direction. The advantage of 
this approach is that the dynamics of the medium are correctly taken into account, as well as the 
geometry of production of the jet pairs. The energy loss is computed via the BDMPS \cite{Dok_etal}
approach. The speed of sound is evaluated locally and the effect of the different flows is taken
into account. The smearing of the back jet is also considered. The results of this computation
can be seen in Figure \ref{RenkRuppert}.  

Finally, in \cite{Friess:2006fk} it was shown that energetic probes propagating through
a strongly coupled $\mathcal{N}=4$ supper Yang Mills plasma emit sound waves. These 
calculations, based on the AdS/CFT correspondence, do not assume thermalization and, even
though  $\mathcal{N}=4$ SYM is not QCD, it is important to understand how this happens in 
this dynamical gauge theory because it may shed light on the jet-medium coupling at RHIC.

\subsection{Transport description}

Without the assumption of thermalization, the backreaction of the medium can be 
addressed via a transport model. Such a study was performed in \cite{Ma:2006fm,Ma:2006mz},
 where the two 
and three particle correlations of the AMPT model were analyzed. This model is 
a $2\leftrightarrow 2$ parton cascade in which the hadronization is performed via
coalescence. The partonic cross sections are tuned to reproduce the
observed elliptic flow. What it is found is that the two particle correlations show
a large angle structure similar to what is expected from the Mach Cone. The 
structure also has a conical shape.  The large angle signal grows with the cross section
and with the system size. All these observations are consistent with the formation 
of sound waves
in 
the transport calculations.

\section{Transfer of Energy to Propagating Modes}
\subsection{Large Angle Gluon Radiation}
The standard mechanism of energy loss, the medium induced gluon radiation,
transfers the energy of the jet into gluons which propagate with the vacuum 
dispersion relations out of the
interaction region and fragment.
The multiple scattering with the medium constituents leads
to the LPM effect which suppresses the zero angle emission and, thus,
the angular distribution of inclusive gluons around the jet axis peaks at an 
angle off zero \cite{Vitev:2005yg,Polosa:2006hb}. For small enough frequency, this angle can be large.

In this case, the delocalization in rapidity of the back jet plays a crucial 
role in the manifestation of this angular distribution in the two particle correlations.
It was noted in \cite{Vitev:2005yg} that the angular dependence of the inclusive distribution is washed out
as a consequence of the smearing. 
However, in the experimental measurements in which the large angle correlations are
observed,  the trigger and associate particles are close in momentum. The tight kinematic window
means that the fragmentation into several gluons is suppressed. This point was noted in \cite{Polosa:2006hb}, 
where
the authors suggested that only the exclusive fragmentation into one gluon is relevant for the 
experimental kinematic region. As a consequence of this, a Sudakov form 
factor is introduced which leads to a much stronger angular dependence of the single gluon radiation probability. 
This strong dependence survives the smearing and the final single gluon distribution peaks at large angles 
similar to those of the experimental correlations. The centrality dependence of the peak position is also 
reproduced.

\subsection{Plasma Modes}

As opposed to the previous case,
for gluons of frequency not much larger than the medium scale
($T\approx 300 MeV$) one may expect that the color 
waves emitted by the high energy particle do not have the vacuum dispersion relations.
%The particle emits plasma modes which may be either longitudinal or transverse.  

In \cite{Koch:2005sx} the authors argued that if the quark gluon plasma at temperatures right above the phase
transition is formed by bound states \cite{Shuryak:2003ty} (as opposed to free quarks and gluons) the dispersion 
relation of the gluons in the medium becomes space like as a consequence of the excitation 
of different levels of those bound states. The frequencies at which this phenomenon happens
are of the order of the interlevel spacing. 

The fact that the dispersion relations become space-like means that there is spontaneous
gluon emission from the probe.
In this frequency region, the in-medium speed of the gluons is smaller
than 1 and ultrarelativistic particles lead to Cherenkov radiation \cite{Koch:2005sx,Dremin:2005an}
 which has 
analogous consequences to the Mach Cone for particle correlations. Unforturnately, a concrete model 
for the structure of the bound states able to explain the experimental correlations was not provided. 

%Let us mention that in this mechanism, for a fixed energy of the trigger particle, it is expected
%that the large angle structure will dissapear as one increases the associate momentum particle momentum.
%This is because  it is only for frequencies of the order of the level spacing where the dispersion 
%relations become space-like. As one increases the frequency the dispersion relation becomes the usual one. 
%Also, since the observed structure occurs at energies of the order of 1-2 Gev, the present mechanism 
%demands the presence of heavy bound states. 

A similar mechanism of conical emission has been advocated in \cite{Ruppert:2005uz},
where it is the plasmon (longitudinal mode) the
one that becomes space like.  

\section{Deflected Jets}
All the previous scenarios are characterized by conical emission of particles
in the final state around the jet axis. However some authors have proposed 
mechanisms in which the large angle structure observed in the dihadron 
correlations is due to a shift of the entire shower of particles coming from
the jet to a finite angle off the jet axis.

In \cite{Armesto:2004pt} it was argued that the strong flow of matter at RHIC modifies the gluon 
emission distributions making them non symmetric around the jet axis. This is
a consequence of the increase  with 
the fluid velocity
of the mean  transferred momentum  to the jet. This idea was originally proposed to explain the rapidity
elongation of the jets, but the strong radial flow can lead to a similar 
effect in the transverse plane. 

Another possibility explored in \cite{Chiu:2006pu} is that intermediate momentum particles receive
large kicks from the particles in the bath as they propagate out the medium. 
These kicks lead to deflection of the direction of propagation of the parton,
which in \cite{Chiu:2006pu} was treated as a Markovian process. The cooling of the medium, which 
leads to a reduction of the typical momentum transfer with time, ensures that 
the final distribution of particles peaks at certain angle with respect to the
jet axis. Estimates of the necessary typical angular deviation of the 
jet after this kicks leads to very large values. 

%The fact that in this scenarios the emission is not conical means that they 
%are clearly distinguishable from the previous by means of three particle 
%correlations.  

\section{Conclusions}
The measured dihadron azimuthal distributions at RHIC have inspired several models
that attempt to explain the observed data. 

The good hydrodynamical behavior of the matter formed in AuAu collisions implies that
the medium reacts collectively to the jet passage, leading to the formation of 
a Mach shock from sound emission \cite{Stocker,Casalderrey-Solana:2006sq,Casalderrey-Solana:2004qm}. 
Unfortunately, the freedom in the coupling of the
high energy partons and the hydro fields does not allow to make a clean prediction of the
observed magnitude of the correlations. Under the assumption that the medium excitation is 
isentropic, only sound is produced and the particle production associated to the jet passage 
is peaked at the Mach angle, only dictated by the speed of sound in the medium,
$\Delta \phi=\pi \pm 1.2 $ rad. which agrees with the experimental observation. For 
a fixed trigger $p_t$, the large angle structure disappears for low $p_t$ associate
particles. As the trigger $p_t$ increases, the contribution of the jets that go through
the medium fills up the dip at $\Delta \phi =\pi$, but the medium contribution remains 
the same for soft particles, in particular it does not change its width. These two effects
have been also observed \cite{Horner}. Finally, the particle emission is conical, which is consistent
with the current three particle correlation analysis \cite{threePPhenix,threePStar}.

The exclusive medium induced one gluon emission distribution also shows features similar 
to the experimental dihadron distributions \cite{Polosa:2006hb}. In this case
the opening angle is solely dependent of the jet energy and the path length traveled.
The centrality dependence of the peak is reproduced.
Even though there is not explicit calculation yet, the splitting angle should decrease 
with the associate $p_t$ for a fixed trigger $p_t$ since  
the kinematic window is larger and the process becomes more exclusive. A similar effect leads to 
fill up of the dip at $\pi$. Finally, only for the configuration in which the gluon and the parton
deflect by a similar angle, this mechanism leads to a conical signal in the 3 particle correlations. 
Even though it is plausible that is the dominant one, an estimate of the magnitude of this configuration with respect to others 
has not yet been done. 

The emission of plasma modes with space like dispersion relations 
\cite{Koch:2005sx,Majumder:2005sw,Dremin:2005an}, different from what 
is obtained by usual perturbative techniques, has been also advocated. In this case, the 
space like character is postulated in order to support the formation of Mach like structures.
A physical mechanism based on the presence of multi level bound states right above the phase transition
was presented in \cite{Koch:2005sx} , which leads to such dispersion relation. However, in order to accommodate the 
experimental observations, heavy bound states ($M\sim 1-2 \,$GeV) are needed. Even though the statement depends 
on the detailed bound state structure, one expects that the opening angle decreases with 
the associate $p_t$, in contradiction with the experimental findings.

Finally, the scenarios based on jet deflection \cite{Chiu:2006pu,Armesto:2004pt} 
seem disfavored by the recent results on three 
particle correlations \cite{threePPhenix,threePStar}.

\ack
Some of the results presented in this talk were worked out with E. Shuryak and D. Teaney. Discussion with 
V. Koch, A. Majumder, T. Renk, J. Ruppert, C. Salgado and X. N. Wang were very helpful in preparing this
talk.

\end{document}